\documentstyle[prl,aps,epsfig,tighten,twocolumn,cite]{revtex}
\begin{document}
\draft
\title{Oscillatory decay of a two-component Bose-Einstein condensate}
\author{Sigmund Kohler$^{1,2}$ and Fernando Sols$^1$}
\address{$^1$Departamento de F\'\i sica Te\'orica de la Materia Condensada
        and Instituto ``Nicol\'as Cabrera'',
        \\Universidad Aut\'onoma de Madrid,
        E-28049 Madrid, Spain}
\address{$^2$Institut f\"ur Physik, Universit\"at Augsburg,
Universit\"atsstr.~1, D-86135 Augsburg, Germany}
\date{\today}
\maketitle
%
\begin{abstract}
We study the decay of a two-component Bose-Einstein condensate
with negative effective interaction energy. With a decreasing atom
number due to losses, the atom-atom interaction becomes less
important and the system undergoes a transition from a bistable
Josephson regime to the monostable Rabi regime, displaying
oscillations in phase and number. We study the equations of motion
and derive an analytical expression for the oscillation amplitude.
A quantum trajectory simulation reveals that the classical
description fails for low emission rates, as expected from
analytical considerations.
Observation of the proposed effect will provide evidence for
negative effective interaction.

\end{abstract}

The experimental realization of Bose-Einstein condensates
\cite{Anderson1995a} offers the possibility of studying quantum
matter waves. A paradigmatic effect in such systems is the
coherent exchange of particles between weakly connected states,
first predicted for superconductors \cite{Josephson1962a}. This
so-called Josephson effect is expected to be displayed also by a
Bose-Einstein condensate (BEC) in a split trap
\cite{Smerzi1997a,Zapata1998a,Marino1999a}. It has been
argued that an ``internal'' Josephson effect may also occur in
condensates whose atoms can reside in two different hyperfine
states coupled by Raman transitions
\cite{Milburn1997a,Steel1998a,Sols1999a,Villain1999a}. Depending
on the relative strength of the interaction between atoms and the
hopping between states, the system may lie in the Rabi regime,
where atoms behave as independent, or in the Josephson regime,
where interactions are sufficiently strong to create a new
collective behavior \cite{Sols1999a}. The coherent exchange of
atoms in the Rabi regime has been realized between two hyperfine
states of $^{87}$Rb atoms \cite{Myatt1997a,Hall1998a,Hall1998b}.

It is possible to have two-component condensates in which it is
energetically favorable for atoms to be in the same state. This
``ferromagnetic'' coupling results from an appropriate combination
of the three relevant scattering lengths \cite{Cirac1998a}, and
does not necessarily require a negative scattering length
\cite{Dalfovo1999a}. On the other hand, atom losses by one- and
three-body collisions cause the condensate to decay. In this
article, we show that the decay of a two-component condensate with
ferromagnetic coupling may display spontaneous oscillations due to
a fundamental dependence of the energy on the relative particle
number.

We consider Bose condensed atoms with two internal hyperfine
levels $|A\rangle$ and $|B\rangle$. The atoms interact via $AA$,
$BB$, and $AB$ elastic collisions. Additionally, a laser field
induces Raman transitions $|A\rangle\leftrightarrow|B\rangle$,
generating an internal Josephson effect
\cite{Sols1999a,Hall1998a}. If the orbital wave functions of
states $|A\rangle$ and $|B\rangle$ may be regarded as rigid, the
system can be adequately described by the two-mode Hamiltonian
\cite{Milburn1997a,Steel1998a,Sols1999a,Cirac1998a}
\begin{equation}
\label{hamiltonian}
H = -\frac{\hbar\omega_R}{2}\left(a^\dagger b + ab^\dagger\right)
    +\frac{E_c}{8} \left(a^\dagger a - b^\dagger b\right)^2,
\end{equation}
where the Rabi frequency $\omega_R$ depends on the detailed laser
tuning, and where the ``charging'' energy
$E_c=(u_{AA}+u_{BB}-2u_{AB})/4$ encompasses the various
interactions. The effective interaction constants are given by
$u_{ij}=(4\pi\hbar^2a_{ij}/m)\int d^3\!x|\varphi_i(x)|^2|
\varphi_j(x)|^2$, where the mode functions $\varphi_{i}(x)$,
$i=A,B$, are normalized to unity. We have assumed $u_{AA}=u_{BB}$,
so that a contribution linear in the atom number difference
vanishes. In the present work we focus on the case where the
effective interaction energy for the internal Josephson dynamics
is negative ($E_c<0$). This is also the case for a double-well
condensate with a negative scattering length.

First we analyze the structure of the classical phase space. For
that in the Heisenberg equations of motion, we replace the
operators $a$,$b$ by $\sqrt{n_A}e^{-i\phi_A}$,
$\sqrt{n_B}e^{-i\phi_B}$. For the variables $z \equiv
(n_A-n_B)/N\equiv 2n/N$ and $\phi \equiv (\phi_A-\phi_B)$ this
yields
\begin{eqnarray}
\label{phi_dot}
\dot \phi &=& \frac{z}{\sqrt{1-z^2}}\cos\phi -\lambda z \label{phidot} , \\
\label{z_dot}
\dot z &=& -\sqrt{1-z^2} \sin\phi \label{zdot} .
\end{eqnarray}
Hereafter, $\lambda\equiv -NE_c/2\hbar\omega_R>0$ gives the
effective interaction strength, and the time is measured in units
of $1/\omega_R$. Equations (\ref{phi_dot}),(\ref{z_dot}) can also
be obtained from the classical nonrigid pendulum Hamiltonian
\cite{Smerzi1997a,Marino1999a,Sols1999a}
\begin{equation}
\label{H_pendulum} H(z,\phi,\lambda)=-\sqrt{1-z^2}\cos\phi -
\frac{\lambda}{2} z^2 ,
\end{equation}
where $(z,\phi)$ are canonically conjugate coordinates.

The pendulum Hamiltonian (\ref{H_pendulum}) obeys the symmetry
$H(z,\phi,\lambda)=-H(z,\phi+\pi,-\lambda)$ and, thus, a sign change
of the effective interaction $\lambda$ corresponds to a phase
shift $\phi\to\phi+\pi$ plus time reversal. This means that the
$\pi$-states which are found for positive $E_c$
\cite{Raghavan1999a} at maxima of the Hamiltonian, correspond to
minima at $\phi=0$ for negative $E_c$.

The classical fixed points of the pendulum Hamiltonian can easily
be found by linearizing around its extrema.  For $0\le \lambda <1$
one finds the stable solution
\begin{equation}
\label{z_0} z_0=0, \quad \phi_0=0,\quad \Omega_0=\sqrt{1-\lambda},
\end{equation}
where $\Omega_0$ is the frequency of small oscillations around the
potential minimum.  For $\lambda>1$ its potential curvature
becomes negative. Thus at $\lambda=1$ there is a bifurcation
whereby $z_0$ becomes unstable and splits into two
symmetry-related fixed points
\begin{equation}
\label{z_plusminus} z_\pm=\pm\sqrt{1-1/\lambda^2}, \quad
\phi_\pm=0, \quad \Omega_\pm=\sqrt{\lambda^2-1}.
\end{equation}
As $\lambda$ is proportional to the atom number, a decaying
condensate will necessarily cross the bifurcation point
$\lambda=1$, a process which we study below
\cite{comment1}.

\begin{figure}
\centerline{ \psfig{width=7.0cm,figure=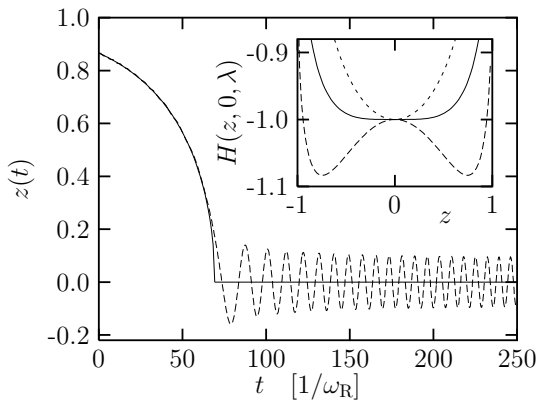} }
\vspace*{3ex} \caption{\label{fig:cl_orbit} Classical time
evolution of the relative number imbalance $z(t)$ (dashed) for
$\gamma=0.01$ and $\lambda_0=2$. The full line denotes the
equilibrium value for the corresponding total atom number. The
inset shows the effective potential (\ref{H_pendulum}) at $\phi=0$
for the interaction strengths $\lambda=1.5$ (dashed), $1$ (full
line), and $0.5$ (dotted). }
\end{figure}%
The bifurcation process can be appreciated in the inset of Fig.\
\ref{fig:cl_orbit}, where $H(z,0,\lambda)$ is represented for
several values of $\lambda$. In the conventional case of $E_c>0$,
it has been noted \cite{Sols1999a} that the point where $\Lambda
\equiv E_c N/2\hbar\omega_R=1$ marks the limit between the Rabi
($\Lambda < 1$) and the Josephson ($\Lambda > 1$) regimes.
Thus we may view the transition from (\ref{z_plusminus})
to (\ref{z_0}) as the negative $E_c$ version of the crossover from
an interaction dominated to a hopping dominated regime. Once
$\lambda < 1$, the dynamics is controlled by hopping and becomes
insensitive to the sign of the interaction.

The master equation that results from the inclusion of random
losses is such that the classical equations of motion
(\ref{phi_dot}),(\ref{z_dot}) remain formally the same, but the
effective interaction constant acquires a time dependence,
\begin{equation}
\label{lambda} \lambda(t) = \lambda_0\, e^{-\gamma t} ,
\end{equation}
due to the exponential decrease of the total atom number. We
integrate the classical equations of motion
(\ref{phidot}),(\ref{zdot}), and (\ref{lambda}) with the initial
condition  $z(0)=z_+(\lambda_0)$, $\phi(0)=0$, and $\lambda_0=2$,
i.e., we start from a minimum of the pendulum Hamiltonian
(\ref{H_pendulum}) in the bistable regime, where most of the atoms
are in the state $|A\rangle$. Figure \ref{fig:cl_orbit} shows the
corresponding time evolution of the number imbalance. As the total
number decreases due to losses, the two minima of the effective
potential merge towards a single minimum at $z=0$ (see inset). The
system evolves adiabatically until $\lambda$ approaches and
crosses the critical value of 1. We will see that a departure from
adiabaticity necessarily occurs when $\lambda$ gets sufficiently
close to 1. After crossing the bistability threshold the system
finds itself distant from the new equilibrium point and begins to
oscillate spontaneously around it. These oscillations initiate in
the crossover region between the Josephson and the Rabi regimes.

The oscillation amplitude can be estimated analytically if one
assumes (i) that outside the transition region $\lambda\approx 1$
the effective potential changes sufficiently slowly and (ii) that
the crossover region is traversed fast enough for the process to
be described as a sudden change in the potential.

For $\lambda > 1$ the condition for adiabaticity reads
\begin{equation}
\label{adiabatic1} 1 >
\left|\frac{d}{dt}\frac{1}{\Omega_\pm}\right|
  = \frac{\gamma\lambda^2}{\Omega_\pm^3},
\end{equation}
while later, for $\lambda < 1$, the condition is
\begin{equation}
\label{adiabatic2} 1 > \left|\frac{d}{dt}\frac{1}{\Omega_0}\right|
= \frac{\gamma\lambda}{2\Omega_0^3}.
\end{equation}
For $\lambda \approx 1$ the r.h.s.\ of the inequalities (\ref{adiabatic1})
and (\ref{adiabatic2}) become $\gamma/[2(\lambda-1)]^{3/2}$ and
$\gamma/2(1-\lambda)^{3/2}$, respectively. This indicates that,
sufficiently close to the bifurcation point $\lambda=1$, adiabaticity breaks
down.

Initially, the system follows the potential minimum
$z_+(\lambda(t))$ adiabatically until the inequality
(\ref{adiabatic1}) is violated for the first time. This occurs for
\begin{equation}
\label{z_init} z=
z_+(\lambda)\Big|_{\Omega_\pm^3(\lambda)=\gamma\lambda^2} \approx
\gamma^{1/3},
\end{equation}
where $\lambda \approx 1$ has been assumed. As $\lambda$ crosses
the transition region, we may assume that there is a sudden change
of the effective potential in the sense that $z$ and $\phi$ are
effectively constant. Then the adiabatic condition, now reading as
in (\ref{adiabatic2}), is recovered for $\lambda$ sufficiently
below 1.

The resulting dynamics can be conveniently studied by noting that,
for a harmonic oscillator, the action $J=2\pi E/\Omega_0$ is an
adiabatic invariant \cite{Arnold1984a}. The energy
$E=\frac{1}{2}(\phi^2+\Omega_0^2 z^2)$ is estimated from Eq.\
(\ref{z_init}) leading to $J=2\pi\Omega_0 z_+^2 = \pi\gamma$. Thus
the instantaneous amplitude reads $[\gamma/\Omega_0(\tau)]^{1/2}$
and the imbalance oscillates like
\begin{equation}
z(\tau) \approx
\sqrt{\gamma/\Omega_0(\tau)}\cos[\Omega_0(\tau)\tau],
\end{equation}
where $\tau$ is the time elapsed after the oscillation sets in,
and $\Omega_0(\tau) = ({1-e^{-\gamma\tau}})^{1/2}$ is the
instantaneous oscillation frequency. For sufficiently small
$\gamma$, $\Omega_0(\tau)=(\gamma\tau)^{1/2}$ which yields, for
the time $\tau_k$ at which the $k$-th maximum of the oscillation
happens, the condition $2\pi k = \Omega_0(\tau_k)\tau_k \approx
\gamma^{1/2}\,\tau_k^{3/2}$. The corresponding amplitudes
therefore read
\begin{equation}
\label{zk,cl} z_k \approx (\gamma^2/2\pi k)^{1/6}.
\end{equation}
We have numerically checked the values of  $z_k$ for $1 \le k \le
5$ in the range $10^{-5}<\gamma<10^{-1}$, and have found good
agreement with this analytical prediction.

The classical treatment above requires a smooth potential, such
that its curvature is basically constant on the scale of the width
$\Delta z$ of the wave packet. While approaching the transition
point $\lambda=1$, the potential becomes flatter and, thus, the
stationary wave packet spreads. At the same time the distance
$|z_+-z_-|$ between the two potential minima decreases. As long as
\begin{equation}
\label{cl.condition} \Delta z < |z_+-z_-|,
\end{equation}
the classical description should be adequate. When inequality
(\ref{cl.condition}) is violated, a quantum description is
necessary. For that we take the variables $z$ and $\phi$ to obey
the commutation relation $[n,e^{i\phi}]=ie^{i\phi}$, which for
small phase uncertainty amounts to $[z,\phi]=2i/N$. For the
Hamiltonian (\ref{H_pendulum}) in the harmonic approximation, this
yields the number r.m.s.\ $(\Delta z)^2=1/N\Omega_+$.

When $\lambda \rightarrow 1^+$, the adiabaticity condition
(\ref{adiabatic1}) reads $[2(\lambda-1)]^{3/2}>\gamma$, while
(\ref{cl.condition}) amounts to $[2(\lambda-1)]^{3/2}>1/4N$. We
expect the classical description of the passage through the
bifurcation to be correct if, as $\lambda$ decreases in the
bistable region, adiabaticity breaks down before classicality does
\cite{comment2} . This is the case if $N\gamma > 1$, i.e., if
losses are sufficiently intense that more than one atom is lost in
a typical Rabi cycle. On the contrary, we expect the classical
predictions to fail for $N\gamma \alt 1$.
A similar conclusion is reached for the regime $\lambda<1$ if we
replace (\ref{cl.condition}) by the condition that the width
$(N\Omega_0)^{-1/2}$ of the ground state wave function is smaller
than the scale $\sqrt{1-\lambda}$ below which the effective potential
behaves as harmonic.
Below we see that these expectations are confirmed by the
numerical simulation.

To analyze the quantum dynamics of a single experiment, we employ
the quantum jump method \cite{Molmer1996a} where the time evolution
over a sufficiently small time interval $\Delta t$ is
simulated as follows. If an atom is lost from condensate $A$
($B$), the wave function changes as $|\psi(t+\Delta t)\rangle \to
Q|\psi(t)\rangle$, $Q=a,b$. These losses occur with the
probabilities $p_Q(t,\Delta t) = \gamma \Delta t
\langle\psi(t)|Q^\dagger Q|\psi(t)\rangle$. If no loss takes
place, the system propagates form $t$ to $t+\Delta t$ with the
effective Hamiltonian $H_{\rm eff} = H -i\hbar\gamma (a^\dagger a
+ b^\dagger b)/2$. Since the propagation with $H_{\rm eff}$ as
well as the atom losses are non-unitary, the wave function has to
be normalized after each time step. Note that in the case of two
identical loss rates the total atom number commutes also with the
effective Hamiltonian and terms in $H_{\rm eff}$ which depend only
on the total atom number can still be neglected.

We start our simulation in the bistable region with $\lambda_0=2$.
Thus the initial state is the ``local ground state'' at the
potential minimum $z_+$ and satisfies condition (\ref{cl.condition}). Here we
make the very reasonable assumption that the parity symmetry is
effectively broken due to the large value of $N_0$, i.e., that any
Schr\"odinger cat state has evolved into a mixture due to tiny
perturbations from the environment.

\begin{figure}
\centerline{ \psfig{width=7.0cm,figure=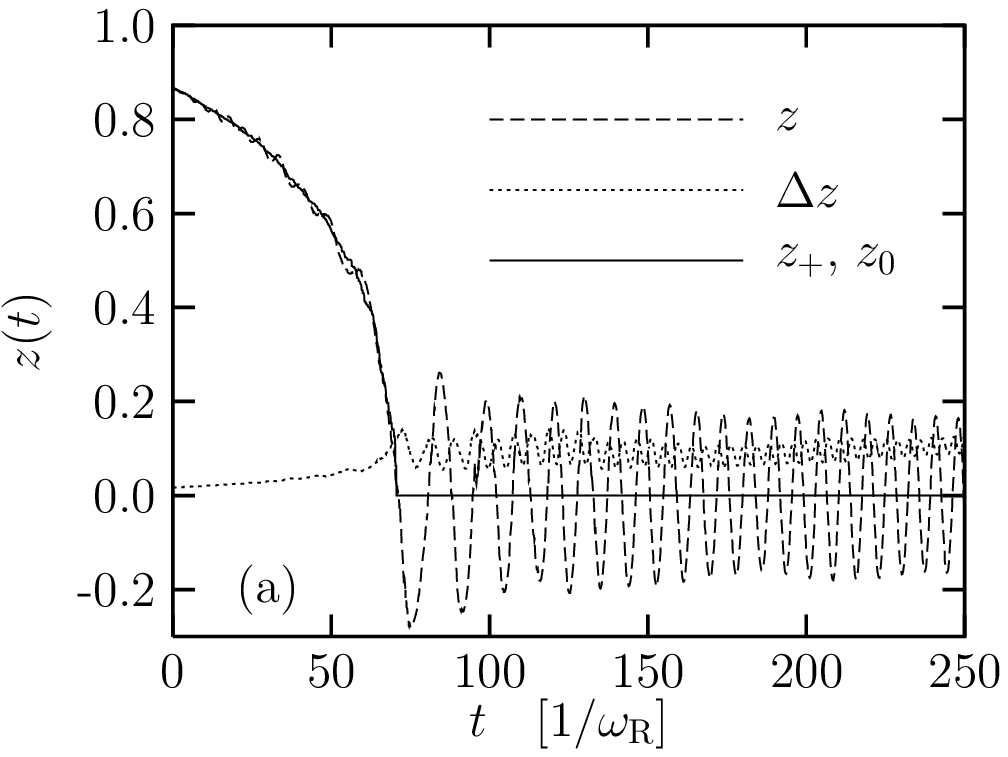} }
\centerline{ \psfig{width=7.0cm,figure=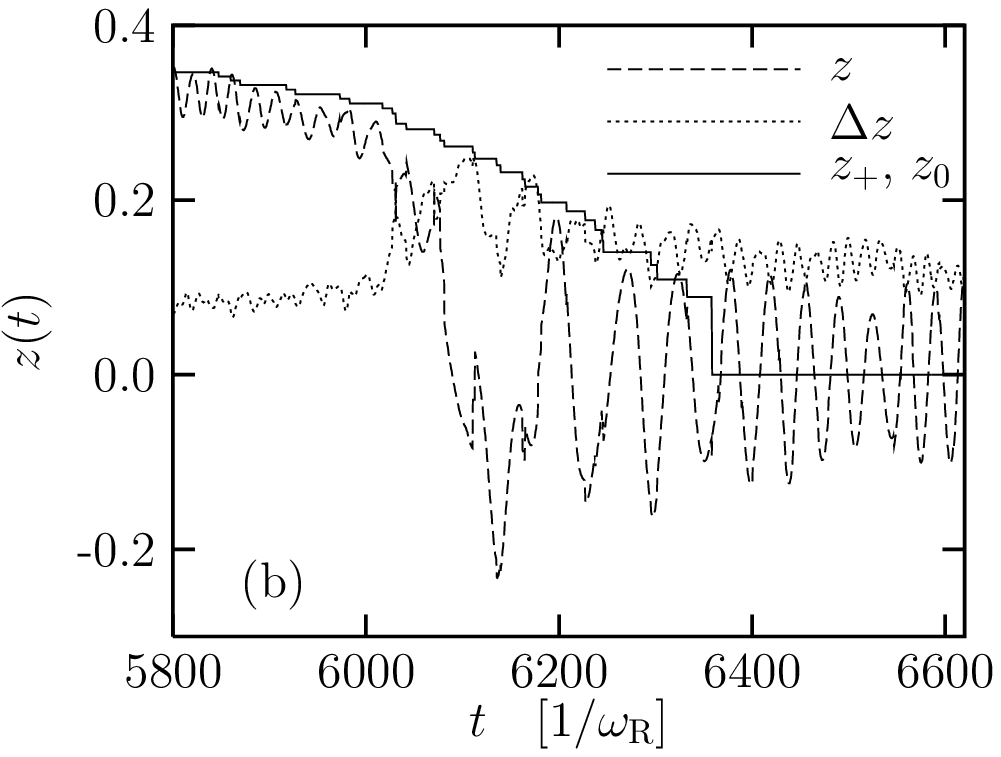} }
\vspace*{3ex} \caption{\label{fig:qm_orbit} Quantum mechanical
time evolution for $N_0=1000$ and $\lambda_0=2$. The decay rate is
$\gamma=10^{-2}$ (a), $10^{-4}$ (b). The steps in the stationary
value $z_+$ are due to the discreteness of the total atom number.
}
\end{figure}

\begin{figure}
\centerline{ \psfig{width=7.0cm,figure=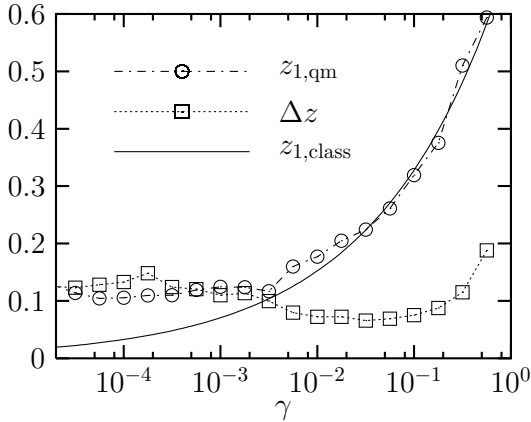} }
\vspace*{3ex} \caption{\label{fig:qm_amplitude} Amplitude of the
quantum mechanical oscillations (circles) and its quantum
uncertainty (boxes) for $N_0=1000$ and $\lambda_0=2$. Each data
point results from an average over 20 simulation runs. The full
line displays the classical result (\ref{zk,cl}) for $k=1$. }
\end{figure}

Figure~\ref{fig:qm_orbit}(a) depicts the quantum mechanical time
evolution for the same parameters as used in
Fig.~\ref{fig:cl_orbit}.  Both plots agree quite well, which
indicates that the system behaves classically. However, for a
smaller value of the decay rate [Fig.~\ref{fig:qm_orbit}(b)] we
find substantial differences: (i) the oscillation starts well
before the transition to the monostable regime and (ii) its
amplitude is larger than in the corresponding classical case (see
Fig.~\ref{fig:qm_amplitude}). This is so because the finite width
of the wave packet allows to cross the point $z=0$ when there is
still a finite barrier.

A comparison of the classical and the quantum mechanical
oscillation amplitude (Fig.~\ref{fig:qm_amplitude}) shows that the
quantum dynamics agrees to the classical dynamics for sufficiently
large values of the decay rate $\gamma$. We find that the quantum
mechanical oscillation amplitude is larger than its classical
counterpart for $\gamma\lesssim 3\times 10^{-3}$, i.e.,
roughly at the value for which the classical description was
predicted to fail. Then the amplitude is comparable to the phase
uncertainty and therefore difficult to observe. A decay rate
$N_0\gamma\approx 10$ seems optimal for the experimental
observation, since then there are still sufficiently many
oscillations before the condensate has decayed and the amplitude
is still larger than the uncertainty.

Let us assume that the rigid wave functions $\varphi_{A,B}(x)$ are
given by the Thomas-Fermi (TF) approximation for 1000 $^{87}$Rb atoms
in an isotropic trap with frequency $100\,{\rm Hz}$.  This results
in a TF radius of $4\,\mu{\rm m}$ and a harmonic oscillator length
of $1\,\mu{\rm m}$. An effective negative scattering length
$a_{\rm eff}\equiv a_{AA}+a_{BB}-2a_{AB} \approx-a_{AA}/100\approx
-0.5\,{\rm nm}$ yields $E_c/\hbar\approx-0.3 \, {\rm s}^{-1}$.
Then, for $N_0=10^3$, to have $\lambda_0=2$ requires a Rabi
frequency $\omega_R\approx 2\pi\cdot10\,{\rm Hz}$, the ``optimum
value'' for the decay rate corresponds to a condensate life time
of $(\gamma\omega_R)^{-1} \sim 1.5\,{\rm s}$.

In summary, we have shown that a two-component condensate with
negative interaction energy ($E_c<0$) will exhibit spontaneous
oscillations in phase and number during its decay due to atom
losses. These oscillations have their root in the fundamental
dependence of the Josephson coupling energy on the relative boson
number. As the atom number decreases, interaction between atoms
becomes less important than hopping between states. Thus the
system evolves from a Josephson regime, which is bistable due to
the negative sign of $E_c$, to a monostable Rabi regime dominated
by hopping. A classical description of the oscillatory decay
dynamics gives a good quantitative account of the decay process
for high enough loss rates. The observation of this effect will
provide unambiguous evidence of the Josephson effect with negative $E_c$,
by identifying the existence of a bifurcation that is manifested
through the spontaneous appearance of Josephson-Rabi oscillations.
By accurately monitoring the decay process, it will be possible to
quantitatively check our current understanding of the internal
Josephson dynamics of two-component condensates.

We would like to thank J.I.~Cirac and P.~Zoller for valuable
discussions. This work has been supported by the EU TMR Programme
under Contract No.\ MRX-CT96-0042, by the Direcci\'on General de
Investigaci\'on Cient\'{\i}fica y T\'ecnica under Grant No.\
PB96-0080-C02, and by the Acci\'on Integrada No. HU1998-0003.



\begin{thebibliography}{10}
\bibitem{comment2}
The rationale for this criterion is that, if quantum mechanics is
important only in the the nonadiabatic region, then it is not
needed, since in that short time interval no dynamics has time to
manifest itself.
\end{thebibliography}
\end{document}